\documentclass[aps,prc,reprint,superscriptaddress]{revtex4-1}
\usepackage{graphicx}
\usepackage{longtable}
\usepackage{epstopdf}
\usepackage{dcolumn}
\usepackage{booktabs}
\usepackage{dcolumn}
\newcolumntype{x}{D{.}{.}{4.4}}
\newcolumntype{y}{D{.}{.}{2.4}}
\newcolumntype{z}{D{.}{.}{3.10}}
\newcolumntype{a}{D{.}{.}{2.10}}
\newcolumntype{e}{D{.}{.}{2.2}}
\newcolumntype{f}{D{.}{.}{3.8}}

\begin{document}


\title{Shell structure of potassium isotopes deduced from their magnetic moments}


\author{J. Papuga}
\email[]{jasna.papuga@fys.kuleuven.be}
\affiliation{KU Leuven, Instituut voor Kern- en Stralingsfysica, 3001 Leuven, Belgium}
\author{M. L. Bissell}
\affiliation{KU Leuven, Instituut voor Kern- en Stralingsfysica, 3001 Leuven, Belgium}
\author{K. Kreim}
\affiliation{Max-Plank-Institut f\"{u}r Kernphysik, D-69117 Heidelberg, Germany}
\author{C. Barbieri}
\affiliation{Department of Physics, University of Surrey, Guildford GU2 7XH, United Kingdom}
\author{K. Blaum}
\affiliation{Max-Plank-Institut f\"{u}r Kernphysik, D-69117 Heidelberg, Germany}
\author{M. De Rydt}
\affiliation{KU Leuven, Instituut voor Kern- en Stralingsfysica, 3001 Leuven, Belgium}
\author{T. Duguet}
\affiliation{CEA-Saclay, IRFU/Service de Physique Nucl\'{e}are, 91191 Gif-sur-Yvette, France}
\affiliation{National Superconducting Cyclotron Laboratory and Department of Physics and Astronomy, Michigan State University, East Lansing, Michigan 48824, USA}
\author{R. F. Garcia Ruiz}
\affiliation{KU Leuven, Instituut voor Kern- en Stralingsfysica, 3001 Leuven, Belgium}
\author{H. Heylen}
\affiliation{KU Leuven, Instituut voor Kern- en Stralingsfysica, 3001 Leuven, Belgium}
\author{M. Kowalska}
\affiliation{Physics Department, CERN, CH-1211 Geneva 23, Switzerland}
\author{R. Neugart}
\affiliation{Institut f\"{u}r Kernchemie, Universit\"{a}t Mainz, D-55128 Mainz, Germany}
\author{G. Neyens}
\affiliation{KU Leuven, Instituut voor Kern- en Stralingsfysica, 3001 Leuven, Belgium}
\author{W. N\"{o}rtersh\"{a}user}
\affiliation{Institut f\"{u}r Kernchemie, Universit\"{a}t Mainz, D-55128 Mainz, Germany}
\affiliation{Institut f\"{u}r Kernphysik, TU Darmstadt, D-64289 Darmstadt, Germany}
\author{M. M. Rajabali}
\affiliation{KU Leuven, Instituut voor Kern- en Stralingsfysica, 3001 Leuven, Belgium}
\author{R. S\'{a}nchez}
\affiliation{GSI Helmholtzzentrum f\"{u}r Schwerionenforschung, D-64291 Darmstadt, Germany}
\affiliation{Helmholtz Institut Mainz, Johannes Gutenberg Universit\"{a}t Mainz 55099 Mainz, Germany}
\author{N. Smirnova}
\affiliation{CENBG (CNRS/IN2P3-Universit\'{e} Bordeaux 1) Chemin du Solarium, BP 120, 33175 Gradignan, France}
\author{V. Som$\grave{\rm{a}}$}
\affiliation{CEA-Saclay, IRFU/Service de Physique Nucl\'{e}are, 91191 Gif-sur-Yvette, France}
\author{D. T. Yordanov}
\affiliation{Max-Plank-Institut f\"{u}r Kernphysik, D-69117 Heidelberg, Germany}
\affiliation{Institut de Physique Nucl\'{e}aire Orsay, IN2P3/CNRS, 91405 Orsay Cedex, France}

\date{\today}

\begin{abstract}
\begin{description}
\item[Background] Ground-state spins and magnetic moments are sensitive to the nuclear wave function, thus they are powerful probes to study the nuclear structure of isotopes far from stability. 
\item[Purpose] Extend our knowledge about the evolution of the $1/2^+$ and $3/2^+$ states for K isotopes beyond the $N = 28$ shell gap.
\item[Method] High-resolution collinear laser spectroscopy on bunched atomic beams.
\item[Results] From measured hyperfine structure spectra of K isotopes, nuclear spins and magnetic moments of the ground states were obtained for isotopes from $N = 19$ up to $N = 32$. In order to draw conclusions about the composition of the wave functions and the occupation of the levels, the experimental data were compared to shell-model calculations using SDPF-NR and SDPF-U effective interactions. In addition, a detailed discussion about the evolution of the gap between proton $1d_{3/2}$ and $2s_{1/2}$ in the shell model and {\it{ab initio}} framework is also presented.   
\item[Conclusions] The dominant component of the wave function for the odd-$A$ isotopes up to $^{45}$K is a $\pi 1d_{3/2}^{-1}$ hole. For $^{47,49}$K, the main component originates from a $\pi 2s_{1/2}^{-1}$ hole configuration and it inverts back to the $\pi 1d_{3/2}^{-1}$ in $^{51}$K. For all even-$A$ isotopes, the dominant configuration arises from a $\pi 1d_{3/2}^{-1}$ hole coupled to a neutron in the $\nu 1f_{7/2}$ or $\nu 2p_{3/2}$ orbitals. Only for $^{48}$K, a significant amount of mixing with $\pi 2s_{1/2}^{-1} \otimes \nu (pf)$ is observed leading to a $I^{\pi}=1^{-}$ ground state. For $^{50}$K, the ground-state spin-parity is $0^-$ with leading configuration $\pi 1d_{3/2}^{-1} \otimes \nu 2p_{3/2}^{-1}$.    
\end{description}
\end{abstract}
     

\maketitle



\section{Introduction}
The shell structure of nuclei established by Goeppert-Mayer \cite{Mayer49} and Haxel \textit{et al.} \cite{Haxel1949} more than 60 years ago is the corner stone of nuclear structure described by the shell model. However, a few decades later, with systematic studies of nuclei with large $N / Z$ ratio, known as "exotic nuclei", it was observed that the original shell gaps are not preserved and "new" shell closures appear \cite{Krucken2011,Sorlin2008,Wienholtz2013}. This fact continues to attract the attention of many experimentalists and theorists who try to understand the origin of these changes. Nowadays, despite the experimental challenges, a large variety of exotic nuclei can be produced and studied with highest precision in facilities around the world \cite{Blaum2013,VanDuppen2011}. These experimental data are used by theorists for fine tuning of the effective interactions in order to improve their descriptive as well as predictive power \cite{Heyde2013}.           

In the past decade, the region below Ca ($Z < 20$) with $ 20 \leq N \leq 28$ was investigated intensively, in particular the evolution of the $\pi sd$ orbitals as a function of neutron number (for review see e.g. Refs.\,\cite{Sorline2013,Gade2006}). The energy spacing between the $1/2^{+}$ and $3/2^{+}$ levels as a function of the $\nu f_{7/2}$ occupancy and the evolution of the $N = 20$ and $N = 28$ shell gaps with decreasing $Z$ for odd-A K ($Z = 19$), Cl ($Z = 17$) and P ($Z = 15$) was presented by Gade \textit{et al.} \cite{Gade2006}, with experimental results compared to shell-model calculations up to $N = 28$. The inversion of the $1/2^{+}$ and $3/2^{+}$ states in the Cl chain is observed for the half-filled $\nu 1f_{7/2}$ orbital. The same effect appears for potassium isotopes, but only when the same orbital is completely filled, at $N = 28$. In addition, the evolution of the effective single-particle energies (ESPE) for potassium isotopes (single-hole states in Ca isotopes) based on shell-model calculations is discussed by Smirnova \textit{et al.} in Ref.~\cite{Smirnova2012}, where a degeneracy of the $\pi 2s_{1/2}$ and $\pi 1d_{3/2}$ levels is predicted to occur at $N = 28$ and returns to a "normal" ordering ($\pi 2s_{1/2}$ below $\pi 1d_{3/2}$) approaching $N = 40$ (Fig\,1(c) in Ref.\,\cite{Smirnova2012}). The reordering of the orbitals is driven by the monopole part of the proton-neutron interaction, which can be decomposed into three components: the central, vector and tensor. Initially Otsuka {\it{et al.}} \cite{Otsuka2005} suggested that the evolution of the ESPEs is mainly due to the tensor component. However, in more recent publications \cite{Smirnova2010,Smirnova2012,Otsuka2010} several authors have shown that both the tensor term as well as the central term have to be considered. 

Regarding the shell model, potassium isotopes are excellent probes for this study, with only one proton less than the magic number $Z = 20$. Nevertheless, little and especially conflicting information is available so far for the neutron-rich potassium isotopes. Level schemes based on the tentatively assigned spins of the ground state were provided for $^{48}$K \cite{Krolas2011} and $^{49}$K \cite{Broda2010}. In addition, an extensive discussion was presented by Gaudefroy \cite{Gaudefroy2010} on the energy levels and configurations of $N = 27, 28$ and 29 isotones in the shell-model framework and compared to the experimental observation, where available. However, the predicted spin of $2^{-}$ for $^{48}$K, is in contradiction with $I^{\pi}=(1^{-})$ proposed by Kr\'{o}las \textit{et al.} \cite{Krolas2011}. In addition, the nuclear spin of the ground state of $^{50}$K was proposed to be $0^{-}$ \cite{Baumann1998} in contrast to the recent $\beta$-decay studies where it was suggested to be $1^{-}$ \cite{Crawford2009}. The ground state spin-parity of $^{49}$K was tentatively assigned to be ($1/2^+$) by Broda {\it{et al.}} \cite{Broda2010}, contrary to the earlier tentative $(3/2^+)$ assignment from $\beta$-decay spectroscopy \cite{Carraz1982}. For $^{51}$K, the nuclear spin was tentatively assigned to be $(3/2^{+})$ by Perrot \textit{et al.} \cite{Perrot2006}. 

\begin{figure}
\includegraphics[width=1.0\linewidth]{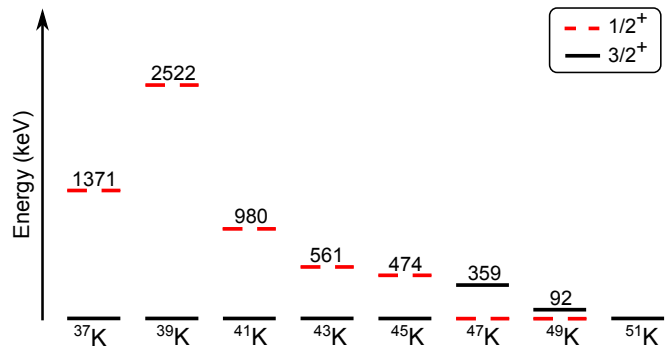}
\caption{\label{fig: Levels-37-51K}(color online) Experimental energies for $1/2^{+}$ and $3/2^+$ states in odd-A K isotopes. Inversion of the nuclear spin is obtained in $^{47,49}$K and reinversion back in $^{51}$K. Results are taken from \cite{Measday2006,Huck1980,Weissman2004,Broda2010}. Ground-state spin for $^{49}$K and $^{51}$K were established \cite{Papuga2013}.} 
\end{figure}

Our recent hyperfine structure measurements of potassium isotopes using the collinear laser spectroscopy technique provided unambiguous spin values for $^{48-51}\rm{K}$ and gave the answer to the question as to what happens with the proton $sd$ orbitals for isotopes beyond $N = 28$. By measuring the nuclear spins of $^{49}\rm{K}$ and $^{51}\rm{K}$ to be $1/2$ and $3/2$ \cite{Papuga2013} respectively, the evolution of these two states in the potassium isotopes is firmly established. This is presented in Fig.\,\ref{fig: Levels-37-51K} for isotopes from $N=18$ up to $N=32$ where the inversion of the states is observed at $N=28$ followed by the reinversion back at $N=32$.  
In addition, we have confirmed a spin-parity $1^{-}$ for $^{48}$K and $0^{-}$ for $^{50}$K \cite{Kreim2014}. The measured magnetic moments of $^{48-51}$K were not discussed in detail so far and will be presented in this article. Additionally, based on the comparison between experimental data and shell-model calculations, the configuration of the ground-state wave functions will be addressed as well. Finally, \textit{ab initio} Gorkov-Green's function calculations of the odd-A isotopes will be discussed.

\section{Experimental procedure}
\begin{figure}
\includegraphics[width=1.0\linewidth]{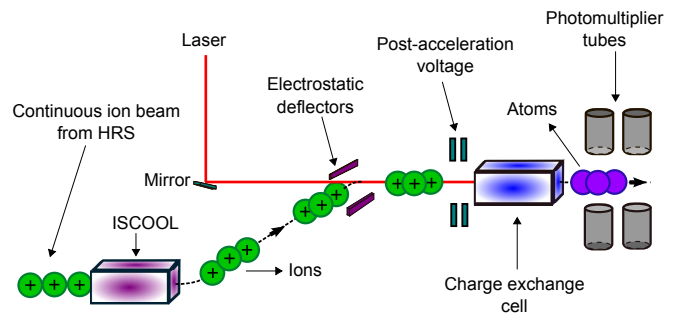}
\caption{\label{beam-line-COLLAPS}(color online) Schematic representation of the setup for collinear laser spectroscopy at ISOLDE.}
\end{figure}

The experiment was performed at the collinear laser spectroscopy beam line COLLAPS \cite{Mueller1983} at ISOLDE/CERN. The radioactive ion beam was produced by 1.4-GeV protons (beam current about 1.7\,$\mu$A) impinging on a thick UC$_{\rm{x}}$ target (45\,g/cm$^{2}$). Ionization of the resulting fragments was achieved by the surface ion source. The target and the ionizing tube were heated to around 2000\,$^{0}$C. The accelerated ions (up to 40\,kV) were mass separated by the high resolution separator (HRS). The gas-filled Paul trap (ISCOOL) \cite{Franberg2008,Vingerhoets2010} was used for cooling and bunching of the ions. Multiple bunches spaced by 90\,ms were generated after each proton pulse. The bunched ions were guided to the setup for collinear laser spectroscopy where they were superimposed with the laser. A schematic representation of the beam line for collinear laser spectroscopy is shown in Fig.~\ref{beam-line-COLLAPS}. 

A cw titanium:sapphire (Ti:Sa) laser was operated close to the Doppler shifted $4s\;{^2S_{1/2}} \rightarrow 4p\;{^2P_{1/2}}$ transition at 769.9\,nm, providing around 1\,mW power into the beam line. Stabilization of the laser system during the experiment was ensured by locking the laser to a reference Fabry-Perot interferometer maintained under vacuum, which in turn was locked to a frequency stabilized helium-neon (HeNe) laser. An applied voltage of $\pm$10\,kV on the charge exchange cell (CEC) provided the Doppler tuning for the ions, which were neutralized through the collisions with potassium vapor. Scanning of the hfs was performed by applying an additional voltage in a range of $\pm$500\,V. The resonance photons were recorded by four photomultiplier tubes (PMT) placed immediately after the CEC. By gating the signal on the PMTs to the fluorescence photons from the bunches, the signal was only recorded for about 6\,$\mu$s when the bunches were in front of the PMTs. Consequently, the background related to the scattered laser light was suppressed by a factor $\sim 10^{4}$ (6\,$\mu$s/90\,ms). More details about the setup can be found in Ref.~\cite{Kreim2014}.

\section{Results}
\begin{figure}
\includegraphics[width=1.0\linewidth]{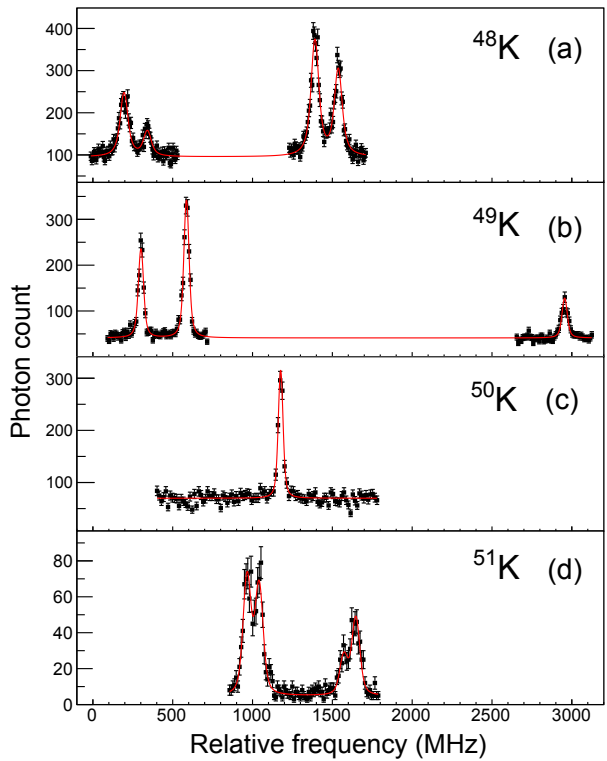}
\caption{\label{Spectra}(color online) The hyperfine spectra of $^{48-51}$K (a-d) obtained by collinear laser spectroscopy. The spectra are shown relative to the centroid of $^{39}$K.}
\end{figure}

In Fig.~\ref{Spectra} typical hyperfine spectra for $^{48-51}$K are shown. 
The raw data are saved as counts versus scanning voltage. The conversion from voltage to frequency was carried out by using the masses from \cite{Wang2012} and applying the relativistic Doppler formula. The spectra were fitted with a Voigt line shape using common width for all components. The $\chi^2$-minimization procedure MINUIT \cite{James1994} was used with $A$-parameters ($A(S_{1/2})$ and $A(P_{1/2})$), the center of gravity and the intensities left as free fit parameters. Nuclear spins, magnetic moments and changes in mean square charge radii were extracted model independently. From the intensity ratios of the hyperfine components, the nuclear spin of $^{48}$K and $^{51}$K were determined to be $I =1$ \cite{Kreim2014} and $I =3/2$ \cite{Papuga2013}, respectively. Since only three peaks are observed in the hyperfine spectrum of $^{49}$K, a spin of $I = 1/2$ can be unambiguously assigned \cite{Papuga2013}. A single peak in the hyperfine spectrum of $^{50}$K corresponds to $I = 0$ \cite{Kreim2014}. The deduced magnetic moments and the implication for the nuclear structure of the potassium isotopes will be reported in this article.

The observed hyperfine $A$-parameters of the ground and the excited states for all studied isotopes are presented in Table~\ref{Results}.
\begin{table*}
\caption{Magnetic hyperfine parameters for neutral potassium from this work and comparison with literature values \cite{Touchard1982,Phillips1965,Chan1969}. \label{Results}}
\begin{ruledtabular}
\begin{tabular}{c c x x x x }
Isotope & 
         \multicolumn{1}{c}{$I^{\pi}$} & 
           \multicolumn{1}{c}{$A(^{2}S_{1/2})$ (MHz)} &
             \multicolumn{1}{c}{$A(^{2}P_{1/2})$ (MHz)} &
               \multicolumn{1}{c}{$A_{\rm{lit}}(^{2}S_{1/2})$ (MHz)} &
                 \multicolumn{1}{c}{$A_{\rm{lit}}(^{2}P_{1/2})$ (MHz)} \\
\hline
$^{38}$K & $3^{+}$ & +404.3\,(3) & +48.9\,(2) & +404.369\,(3) & - \\
$^{38 \rm m}$K & $0^{+}$ & 0 & 0 & - & - \\
$^{39}$K & $3/2^{+}$ & +231.0\,(3) & +27.8\,(2) & +231.0\,(3) & +27.5\,(4) \\
$^{42}$K & $2^{-}$ & -503.7\,(3) & -61.2\,(2) & -503.550779\,(5) & -60.6\,(16) \\
$^{44}$K & $2^{-}$ & -378.9\,(4) & -45.8\,(2) & -378.1\,(11) & -44.9\,(11) \\
$^{46}$K & $2^{-}$ & -462.8\,(3) & -55.9\,(2) & -465.1\,(12) & -55.7\,(13) \\
$^{47}$K & $1/2^{+}$ & +3413.2\,(3)\footnotemark[1] & +411.8\,(2) & +3420.2\,(29) & +411.9\,(50) \\
$^{48}$K & $1^{-}$ & -795.9\,(3) & -96.3\,(3) & - & - \\
$^{49}$K & $1/2^{+}$ & +2368.2\,(14) & +285.6\,(7) & - & - \\
$^{50}$K & $0^{-}$ & 0 & 0 & - & - \\
$^{51}$K & $3/2^{+}$ & +302.5\,(13) & +36.6\,(9) & - & - \\
\end{tabular}
\end{ruledtabular}
\footnotetext[1]{After reanalysis, the uncertainty on this value was increased from 0.2 to 0.3\,MHz.}
\end{table*}
The results are compared to the literature values from \cite{Phillips1965,Touchard1982,Chan1969}. Compared to the results from earlier atomic beam laser spectroscopy studies \cite{Touchard1982}, the precision has been increased by an order of magnitude for most of the values. The hyperfine $A$-parameters for $^{48-51}$K were measured for the first time. For the isotopes$/$isomer with $I = 0$, there is no hyperfine splitting of the atomic states, thus the $A$-parameters are equal to 0.      

The relation between $A$-parameters and magnetic moments is given by: $A = \mu B_{0} / I J$, where $B_{0}$ is the magnetic field induced by the electron cloud at the position of the nucleus. As $B_{0}$ is to first order isotope independent, magnetic moments were deduced relative to $^{39}$K using Eq.~(\ref{eq:Relative_mm}):
\begin{equation}
\label{eq:Relative_mm}
\mu = \frac{A(^{2}S_{1/2})I}{A_{\rm{ref}}(^{2}S_{1/2})I_{\rm{ref}}}\mu_{\rm{ref}}. 
\end{equation} 
   
The reference values were taken from atomic-beam magnetic resonance measurements, where precise values are reported to be $A_{\rm{ref}}(^{2}S_{1/2})= + 230.8598601(7)$\,MHz and $\mu_{\rm{ref}}=+0.3914662(3)\,\mu_{N}$ \cite{Beckmann1974}. 
 
As the magnetic moments of potassium isotopes were determined with $10^{-3}$\,-\,$10^{-4}$ relative precision, one can not neglect the hyperfine structure (hfs) anomaly  between two isotopes, arising from the finite size of the nuclei. This slightly modifies the $A$-parameters \cite{Bohr1951} and gives a small correction of Eq.~(\ref{eq:Relative_mm}) which is expressed by 
\begin {equation}
^{39}\Delta^{\rm{A}} = \frac{A^{39}(S_{1/2})/g(^{39}\rm{K})}{A^{\rm{A}}(S_{1/2})/g(^{A}\rm{K})} - 1
 ,
\label{eq: Def-HFA}
\end {equation}
being different from zero. In Eq.~(\ref{eq: Def-HFA}), the $g$ factor is $g = \mu/I$. The dominant contributions to the hfs anomaly are originating from the difference in the nuclear magnetization distribution (Bohr-Weisskopff effect \cite{BohrWeisskopf1950}) and difference of the charge distribution (Breit-Rosenthal effect \cite{Rosenthal1932}). In the case of potassium isotopes, the hfs anomaly was measured for $^{38-42}$K relative to $^{39}$K \cite{Phillips1965,Eisinger1952,Beckmann1974,Chan1969,Ochs1950}. In order to assess the additional uncertainty on the magnetic moments for all measured isotopes, the hfs anomaly was estimated from the experimental data as well as from theoretical calculations. 

According to the approach proposed by Ehlers {\it{et al.}} \cite{Ehlers1968}, the differential hyperfine structure anomaly ($^{39}\delta^{\rm{A}}$) between two different electronic states is defined as: 

\begin {equation}
^{39}\delta^{\rm{A}} = \frac{A^{39}(S_{1/2})/A^{39}(P_{1/2})}{A^{\rm{A}}(S_{1/2})/A^{\rm{A}}(P_{1/2})} - 1 ,
\label{eq: Exp-HFA}
\end {equation}
where the $A$-parameters for the reference isotope $^{39}$K were taken from literature \cite{Beckmann1974, Falke2006}. The value of the hyperfine structure anomaly can be approximated by the differential hyperfine structure anomaly, which is good to a few percent. This is good enough considering the accuracies of our experimental results. Differential hyperfine anomalies are presented in Table~\ref{HF_anomaly} (col. 3). For $^{40,41}$K, the experimental results from literature were used: the $A(S_{1/2})$ parameter from \cite{Eisinger1952,Beckmann1974}, while the $A(P_{1/2})$ parameters were taken from \cite{Falke2006}. It should be noted that for $^{43,45}$K no data for $A(P_{1/2})$ were obtained. In addition, theoretical calculations were performed following Bohr \cite{Bohr1951}. The hfs anomaly was estimated to be $^{39}\Delta^{\rm{A}}_{\rm{theo}} = \epsilon (^{39}\rm{K})- \epsilon(^{\rm{A}}\rm{K})$, where $\epsilon (^A\rm{K})$ is a perturbation factor due to the finite size of the nucleus. It can be calculated using \cite{Bohr1951}:

\begin {equation}
\epsilon=-[(1+0.38\zeta)\alpha_{s}+0.62\alpha_{l}]b(Z, R_{0})(R/R_{0})^2.
\label{eq: Epsilon-HFA}
\end {equation} 

In Ref.~\cite{Bohr1951}, all parameters from Eq.~(\ref{eq: Epsilon-HFA}) are defined and for some of them values are tabulated. Theoretical estimations of the $\epsilon$ parameter and hfs anomaly ($^{39}\Delta^{\rm{A}}_{\rm{theo}}$) are listed in Table~\ref{HF_anomaly} (col. 4 and 5). Hyperfine structure anomalies of the potassium isotopes known from the literature \cite{Ochs1950,Beckmann1974,Eisinger1952,Chan1969,Phillips1965} are shown in the last column of Table~\ref{HF_anomaly} ($^{39}\Delta_{\rm{lit}}^{\rm{A}}$).
\begin{table}
\caption {Estimated hyperfine structure anomalies of potassium isotopes. Experimental results for the hyperfine parameters were used to calculate ($^{39}\delta^{\rm{A}}$) from Eq.\,(\ref{eq: Exp-HFA}). For $^{40,41}$K experimental data were taken from \cite{Eisinger1952,Beckmann1974,Falke2006}. The $\epsilon (^A\rm{K})$ parameters for all isotopes are calculated from Eq.\,(\ref{eq: Epsilon-HFA}) and are listed in the next column. For the reference isotope, it was found to be $\epsilon (^{39}\rm{K}) = 0.165$. The estimated hyperfine structure anomalies from the model ($^{39}\Delta_{\rm{theo}}^{\rm{A}}$) described by Bohr (see text for details) are shown as well. In the last column, the hyperfine structure anomalies from literature ($^{39}\Delta_{\rm{lit}}^{\rm{A}}$) are given \cite{Ochs1950,Beckmann1974,Eisinger1952,Chan1969,Phillips1965}. 
\label{HF_anomaly}}
\begin{ruledtabular}
\begin{tabular}{c c c c c c}
Isotope & 
  \multicolumn{1}{c}{$I^{\pi}$} & 
    \multicolumn{1}{c}{$^{39}\delta^{\rm{A}}$ (\%)} &       
       \multicolumn{1}{c}{$\epsilon (^A\rm{K})$} & 
          \multicolumn{1}{c}{$^{39}\Delta_{\rm{theo}}^{\rm{A}}$ (\%)}  &
             \multicolumn{1}{c}{$^{39}\Delta_{\rm{lit}}^{\rm{A}}$ (\%)} \\
\hline
$^{38}$K & $3^{+}$ & 0.53\,(44) & -0.006 & 0.17 & 0.17\,(6) \\
$^{40}$K & $4^{-}$ & 0.43\,(17) & -0.379 & 0.54 & 0.466\,(19) \\
$^{41}$K & $3/2^{+}$ & -0.23\,(31) & 0.398 & -0.23 & -0.226\,(10) \\
         &           &             &       &       & -0.22936\,(14) \\
$^{42}$K & $2^{-}$ & 0.99\,(36) & -0.265 & 0.43 & 0.336\,(38) \\
$^{43}$K & $3/2^{+}$ & - & 0.560 & -0.39 & -\\
$^{44}$K & $2^{-}$ & 0.47\,(47) & -0.302 & 0.47 & - \\
$^{45}$K & $3/2^{+}$ & - & 0.521 & -0.36 & - \\
$^{46}$K & $2^{-}$ & 0.40\,(39) & -0.275 & 0.44 & - \\
$^{47}$K & $1/2^{+}$ & 0.28\,(16) & -0.126 & 0.29 & - \\
$^{48}$K & $1^{-}$ & 0.57\,(35) & -0.211 & 0.38 & - \\ 
$^{49}$K & $1/2^{+}$ & 0.24\,(29) & -0.121 & 0.29 & - \\ 
$^{51}$K & $3/2^{+}$ & 0.57\,(250) & 0.097 & 0.07 & - \\ 
\end{tabular}
\end{ruledtabular}
\end{table}   
For all isotopes except $^{42}$K, the hyperfine structure anomaly estimated from the experimental results is in agreement with the calculated ones. The values for odd-odd nuclei are systematically higher than for odd-even, thus we will quote different additional uncertainties on the magnetic moments (in square brackets in Table~\ref{mm-odd-A-Tab} and Table~\ref{mm-even-A}), namely 0.3\% and 0.5\% for odd-A and even-A isotopes, respectively.  
\section{Discussion}
Nuclei with one particle or one hole next to a shell closure are excellent probes for testing shell-model interactions. In this context, the investigation of the potassium chain is of great interest, since it has a hole in the $\pi sd$ orbital and it covers two major neutron shells, $N = 20$ and $N = 28$, and one sub-shell at $N = 32$. 

In what follows, the experimental results from our work are compared to shell-model predictions. The calculations were carried out using the ANTOINE code \cite{Nowacki1999} for two effective interactions: SDPF-NR \cite{Retamosa1997, Nummela2001-ANT} and SDPF-U \cite{Nowacki2009}. The latter is a more recent version of the SDPF-NR interaction where the monopole part was refitted by including more experimental results from nuclei with one particle or one hole next to the closed shell for protons or neutrons such as $^{35}$Si, $^{47}$Ar and $^{41}$Ca. The calculations have been performed in the $0\hbar \omega$ shell model space beyond a $^{16}$O core and with valence protons restricted to $sd$ orbitals and neutrons to $sd$ or $pf$ orbitals. Neutron excitations across $N = 20$ were prohibited. In order to account for missing interactions among the valence nucleons as well as with the nucleons from the core, the calculations were performed using effective $g$ factors: the spin $g$ factors were fixed at $g_{s}^{\rm{eff}}=0.85 g_{s}^{\rm{free}}$, while the orbital $g$ factors were fixed to $g_{l}^{\pi} = 1.15$ and $g_{l}^{\nu}$ = -0.15 \cite{Richter2008}.          

\subsection{Odd-A}

Nuclear properties such as the ground-state spin and magnetic moment of odd-A K isotopes (odd-even isotopes) are determined by an unpaired proton placed in the $\pi sd$ orbital whilst the even number of neutrons are coupled to spin zero. In the simple shell-model framework the measured nuclear spin indicates the dominant component of the ground-state wave function. Based on this simple model, one would expect that the magnetic moments of these isotopes are equal to the single-particle magnetic moments of the orbital where a valence proton is located. However, the observed deviation from the single-particle values reveals influence of the proton-neutron interaction leading to a more collective behavior. Although the magnetic moments of the neutron-rich odd-A K isotopes were already published in \cite{Papuga2013}, a detailed discussion over the entire odd-A chain from $N = 20$ up to $N = 32$ will be presented here with additional focus on the monopole interaction responsible for the shell evolution.

The experimentally observed magnetic moments are listed in Table~\ref{mm-odd-A-Tab} together with the values predicted by shell-model calculations using the SDPF-NR and SDPF-U effective interactions. 
\begin{table*}
\caption{Experimental magnetic moments (in units of $\mu_{\rm{N}}$) compared with the calculated ones using two effective interactions: SDPF-NR and SDPF-U. The predicted amount of the $\pi 1d_{3/2}^{-1}\otimes \nu (fp)$ in the ground-state wave function is given in \%. If available, the literature values are shown as well. The uncertainty in the square brackets is due to the hyperfine structure anomaly  and is 0.3\%. \label{mm-odd-A-Tab}}
\begin{ruledtabular}
\begin{tabular}{c c a c c c c z c}
Isotope & 
\multicolumn{1}{c}{$I^{\pi}$} & 
\multicolumn{1}{c}{$\mu_{\rm exp}$} &
\multicolumn{1}{c}{$\mu_{\rm{SDPF-NR}}$} & 
\multicolumn{1}{c}{$\pi 1d_{3/2}^{-1}$ (\%)} &
\multicolumn{1}{c}{$\mu_{\rm{SDPF-U}}$} &
\multicolumn{1}{c}{$\pi 1d_{3/2}^{-1}$ (\%)} &
\multicolumn{1}{c}{$\mu_{\rm lit}$} & 
\multicolumn{1}{c}{Reference} \\
\hline
$^{39}$K & $3/2^{+}$ & +0.3917\,(5)\,[12] & +0.65 & 100\% & +0.65 & 100\% & +0.3914662\,(3) & \cite{Beckmann1974} \\
$^{41}$K & $3/2^{+}$ & - & +0.37 & 95\% & +0.33 & 95\% & +0.2148701\,(2) & \cite{Beckmann1974} \\
$^{43}$K & $3/2^{+}$ & - & +0.22 & 92\% & +0.17 & 92\% & +0.1633\,(8)\footnotemark[1] & \cite{Touchard1982} \\
$^{45}$K & $3/2^{+}$ & - & +0.23 & 88\% & +0.21 & 90\% & +0.1734\,(8)\footnotemark[1] & \cite{Touchard1982} \\
$^{47}$K & $1/2^{+}$ & +1.9292\,(2)\,[58] & +1.87 & 13\% & +1.91 & 13\% & +1.933(9)\footnotemark[1] & \cite{Touchard1982} \\
$^{49}$K & $1/2^{+}$ & +1.3386\,(8)\,[40] & +1.61 & 21\% & +1.81 & 15\% & - & - \\
$^{51}$K & $3/2^{+}$ & +0.5129\,(22)\,[15] & +0.60 & 90\%  & +0.65 & 93\% & - & - \\
\end{tabular}
\end{ruledtabular}
\footnotetext[1]{Included 0.5\% uncertainty on the error to account for the hyperfine structure anomaly.}
\end{table*}
In the same table, the calculated percentage of the component of the ground-state wave function originating from a hole in the $\pi 1d_{3/2}^{-1}$ is shown as well. 

\begin{figure}
\includegraphics[width=1.0\linewidth]{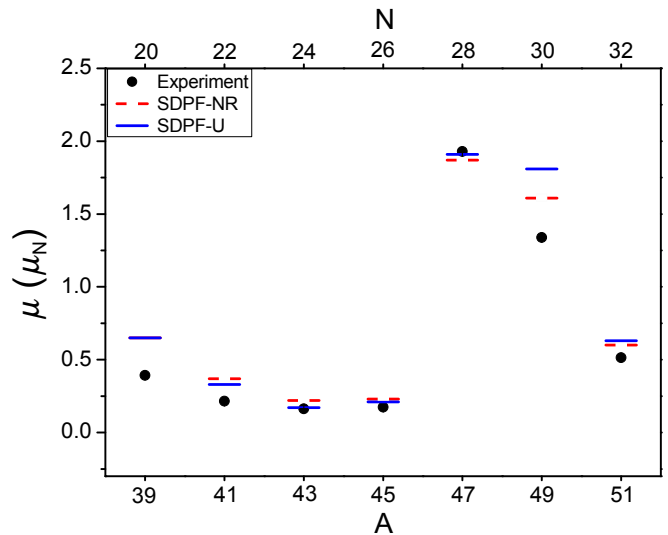}
\caption{\label{Fig-mm-odd-A} (color online) Experimental magnetic moments (black dots) compared to the shell-model calculation using SDPF-NR (red dashed line) and SDPF-U (blue solid line) interactions and effective $g$ factors (see text for more details). In general a very good agreement between experimental and theoretical results is observed, except for $^{39}$K and $^{49}$K.} 
\end{figure}

In Fig.~\ref{Fig-mm-odd-A} the experimental magnetic moments for odd-A K isotopes are compared to the results from the shell-model calculations. 
In general a very good agreement between experimental and theoretical results is observed. The discrepancy for $^{39}$K and $^{41}$K might be due to excitation across the $Z,N=20$ shell gaps, which were not considered in these calculations. This problem is especially pronounced for $^{39}$K where, the shell-model calculations yield a pure $\pi 1d_{3/2}^{-1}$ state with a magnetic moment about $60$\% larger than the experimental value. Both effective interactions yield almost identical amounts of the $\pi 1d_{3/2}^{-1}\otimes \nu (pf)$ component in the ground state of odd-A isotopes. It is more than 90\% for all isotopes up to $^{45}$K, but for $^{47,49}$K the wave function is dominated by the $\pi 2s_{1/2}^{-1} \otimes \nu (pf)$ configuration. This was already concluded in \cite{Papuga2013}, the conclusion  based on the measured ground-state spin and $g$ factor. The only noticeable difference between both calculations is found for $^{49}$K, where the contribution from $\pi 1d_{3/2}^{-1} \otimes \nu (pf)$ is predicted to be 21\% from SDPF-NR and 15\% from SDPF-U. In both cases the calculated value deviates from the experimental one, but SDPF-U shows a larger deviation. From a two-state mixing calculation, at least 25\,\% \cite{Neyens2013} of mixing with the $[\pi 1d_{3/2}^{-1} \otimes \nu (pf)_{2^+}]_{1/2^{+}}$ is needed to reproduce the observed magnetic moment.

The inversion of the nuclear spin from $I = 3/2$ to $I = 1/2$ at $N = 28$ and the re-inversion back to $I=3/2$ at $N=32$ is related to the evolution of the proton orbitals ($\pi sd$) while different neutron orbitals are being filled. This evolution is driven by the monopole term of the nucleon-nucleon ($NN$) interaction. According to Otsuka {\it{et al.}} \cite{Otsuka2010}, the interaction has a linear dependence on the occupation number and consists of three parts: central, vector and tensor. Applying the spin-tensor decomposition method \cite{Smirnova2010,Smirnova2012}, it is possible to separate the contribution of different components of the effective $NN$ interaction. This leads to a qualitative analysis of the role of each part separately in the evolution of the effective single-particle energies (ESPEs). The calculated centroids for every component of the monopole interaction are listed in Table\,\ref{tab: Centroids-SDPF-NR-U}. 
\begin{table*}
\centering
\caption {Spin-tensor content of the centroids of the SDPF-NR and SDPF-U interaction, defining the proton $1d_{3/2}-2s_{1/2}$ gap. Results are presented in MeV.\label{tab: Centroids-SDPF-NR-U}}
\begin{ruledtabular}
\begin{tabular}{c c x x x x x x}
Interaction & 
   Component & 
     \multicolumn{1}{c}{$V_{d_{3/2}f_{7/2}}^{\pi \nu}$} & 
        \multicolumn{1}{c}{$V_{s_{1/2}f_{7/2}}^{\pi \nu}$} & 
           \multicolumn{1}{c}{$\Delta V$} & 
              \multicolumn{1}{c}{$V_{d_{3/2}p_{3/2}}^{\pi \nu}$} & 
                 \multicolumn{1}{c}{$V_{s_{1/2}p_{3/2}}^{\pi \nu}$} & 
                    \multicolumn{1}{c}{$\Delta V$} \\
\hline
\space & Central & -1.66 & -1.26 & -0.40 & -1.34 & -1.46 & +0.12\\
SDPF-NR & Vector & +0.28 & +0.17 & +0.11 & +0.21 & +0.22 & -0.01\\
\space & Tensor & -0.28 & 0.00 & -0.28 & -0.08 & 0.00 & -0.08\\
\cline{2-8}
\space & Total & -1.66 & -1.09 & -0.57 & -1.21 & -1.24 & +0.03 \\
\hline
\space & Central & -1.51 & -1.21 & -0.30 & -1.05 & -1.21 & +0.16 \\
SDPF-U & Vector & +0.09 & +0.07 & +0.02 & +0.05 & -0.11 & +0.16 \\
\space & Tensor & -0.28 & 0.00 & -0.28 &  -0.06 & 0.00 & -0.06\\
\cline{2-8}
\space & Total & -1.70 & -1.14 & -0.56 & -1.06 & -1.32 & +0.26 \\
\end{tabular}
\end{ruledtabular}
\end{table*}
The centroid of the proton-neutron interaction is defined as \cite{Poves1981}:
\begin{equation}
V_{j_\pi j_\nu}=\frac{\sum_{J}(2J+1){\langle j_\pi j_\nu|V|j_\pi j_\nu\rangle}}{\sum_{J}(2J+1)},
\end{equation}
where $j_\pi$ and $j_\nu$ stand for the angular momentum of proton and neutron orbitals, $\langle j_\pi j_\nu|V|j_\pi j_\nu \rangle$ is the two-body matrix element and $J$ is the total angular momentum of a proton-neutron state. The summation runs over all possible values of $J$. 

Based on the results presented in Table\,\ref{tab: Centroids-SDPF-NR-U}, the central component of the interaction is by far the largest (col. 3-4 and 6-7) and, thus has the strongest influence on the energy shift. Note that there is no tensor component for the $s_{1/2}$ orbital due to the absence of a preferred orientation of the spin for an $l = 0$ state \cite{Otsuka2005}. 

\begin{figure}[h!]
\includegraphics[width=0.9\linewidth]{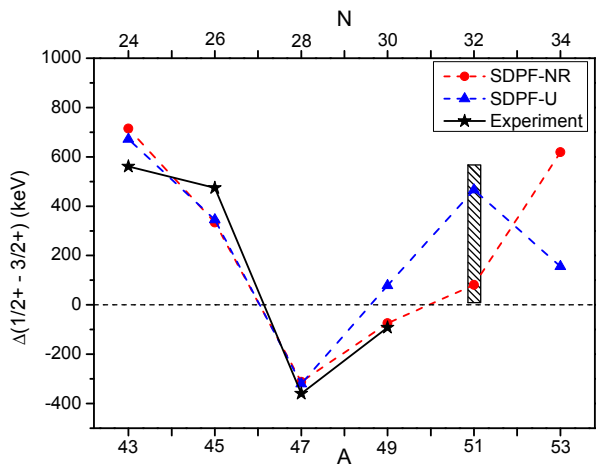}
\caption{\label{Energy-diff} (color online) Energy difference between the two lowest states with $I^{\pi}=1/2^{+}$ and $3/2^{+}$ for odd-A K isotopes from $N = 24$ up to $N = 34$. Experimental results (black stars) taken from \cite{Measday2006,Huck1980,Weissman2004,Broda2010} are in good agreement with the shell-model calculations using different effective interactions: SDPF-NR (red dots) and SDPF-U (blue triangles). For $^{49}$K, only the SDPF-NR interaction correctly predicts the spin of the ground state to be $1/2^{+}$. The shaded area represents the expected region based on the measured ground-state spin and the shell-model calculation for the first excited state in $^{51}$K.   
 }
\end{figure}

The change of the energy gap between $\pi 1d_{3/2}$ and $\pi 2s_{1/2}$ depends on the difference $\Delta V$ between the two centroids (Table\,\ref{tab: Centroids-SDPF-NR-U}; col. 5 and 8). The evolution of the energy gap from $N=20$ to $N=28$ and from $N=28$ to $N=32$, along with the spin-tensor decomposition of this energy gap, is presented in Table\,\ref{tab: ST-decomposition}. 
\begin{table}
\caption {Calculated contributions of the different spin-tensor terms of SDPF-NR ("NR") and SDPF-U ("U") to the evolution of the energy gap between effective $\pi 1d_{3/2}$ and $\pi 2s_{1/2}$ when filling $\nu 1f_{7/2}$ and $\nu 2p_{3/2}$ orbitals. The results are given in MeV. 
\label{tab: ST-decomposition}}
\begin{ruledtabular}
\begin{tabular}{c  c  c  c  c}
filling & 
  \multicolumn{2}{c}{$\nu 1f_{7/2}$} & 
      \multicolumn{2}{c}{$ \nu 2p_{3/2}$} \\
\hline
\space & \rm{NR} & \rm{U} & \rm{NR} & \rm{U} \\
\hline
Central & -2.09 & -1.58 & +0.46 & +0.58 \\
Vector  & +0.58 & +0.06 & -0.06 & +0.43 \\
Tensor  & -1.64 & -1.64 & -0.17 & -0.12 \\
\hline
Total  & -3.15 & -3.16 & +0.23 & +0.89 \\ 
\end{tabular}
\end{ruledtabular}
\end{table}  
Both interactions predict the same decrease of the gap by $-3.15\,\rm{MeV}$ for isotopes from $N = 20$ up to $N = 28$ (Table\,\ref{tab: ST-decomposition}; col. 2-3), although the central and vector contribution are significantly different in both interactions. Once the $\nu p_{3/2}$ orbital is involved, for isotopes from $N=29$ up to $N=32$, the situation changes. The increase in the gap between $\pi 1d_{3/2}$ and $\pi 2s_{1/2}$ (Table\,\ref{tab: ST-decomposition}; col. 4-5) is mostly driven by the central component in the SDPF-NR interaction, while also the vector component contributes significantly in the SDPF-U. Therefore, the calculated change in the energy gap is very different: +0.23\,MeV and +0.89\,MeV, respectively. This results in different calculated spectra for $^{49}$K and $^{51}$K as illustrated in Fig.~\ref{Energy-diff}. This figure shows the energy difference between the lowest $1/2^{+}$ and $3/2^{+}$ states for isotopes in the range from $N = 24$ up to $N = 34$ compared to the calculated values.
Up to $N= 28$, both interactions are in agreement with the experimental results. The deviation between both effective interactions increases beyond $N = 28$ when the $\nu 2p_{3/2}$ and higher orbitals are involved. For $^{49}$K, both interactions calculate the energy difference between the ground and first excited state to be about 75\,keV, but only the SDPF-NR predicts the correct ground-state spin. Although both effective interactions predicted the correct ground-state spin for $^{51}$K, experimental data on the energy of the first-excited state is needed to further test the validity of both models. Beyond $N=32$ the predicted ground-state spin 3/2 for $^{53}$K needs experimental verification, as well as the energy of the first excited $1/2^+$ state, which is very different in both calculations.               
     
\begin{figure}[h!]
\includegraphics[width=0.9\linewidth]{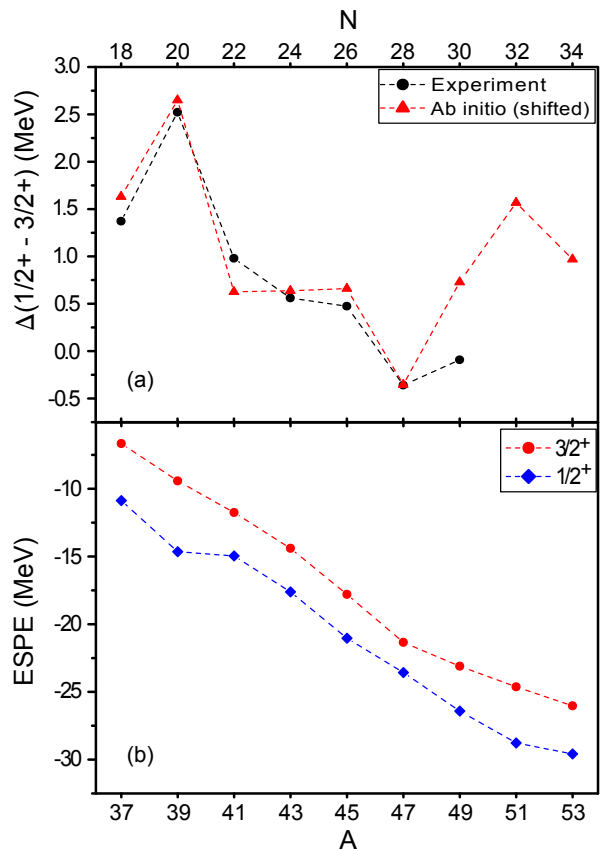}
\caption{\label{AI-ESPE} (color online) (a) Energy difference between the lowest $1/2^{+}$ and $3/2^{+}$ states obtained in $^{37-53}$K from {\it{ab initio}} Gorkov-Green`s function calculations and experiment. {\it{Ab initio}} results have been shifted by 2.58\,MeV to match the experimental $(1/2^{+}-3/2^{+})$ splitting in $^{47}$K. (b) $\pi d_{3/2}$ and $\pi s_{1/2}$ effective single-particle energies (ESPE) in $^{37-53}$K calculated in Gorkov-Green's functions theory.
 }
\end{figure}

Very recently, {\it{ab initio}} calculations of open-shell nuclei have become possible in the Ca region~\cite{Soma2014} on the basis of the self-consistent Gorkov-Green`s function formalism~\cite{Soma2011}. State-of-the-art chiral two- ($NN$)~\cite{Entem2003,Machleidt2011} and three-nucleon ($3N$)~\cite{Navratil2007} interactions adjusted to two-, three- and four-body observables (up to $^4$He) are employed, without any further modification, in the computation of systems containing several tens of nucleons. We refer to Ref.~\cite{Soma2014} for further details. In the present study, Gorkov-Green's function calculations of the lowest $1/2^{+}$ and $3/2^{+}$ states in $^{37-53}$K have been performed by removing a proton from $^{38-54}$Ca. Similarly to Fig.~\ref{Energy-diff}, the upper panel of Fig.~\ref{AI-ESPE} compares the results to experimental data. The calculated energy differences have been shifted down by 2.58 MeV to match the experimental value for $^{47}$K. The overestimation of energy differences is a general feature of calculated odd-A spectra and actually correlates with the systematic overbinding of neighboring even-A ground states \cite{Soma2014}. Still, one observes the correct {\it relative} evolution of the $1/2^{+}$ state with respect to the $3/2^{+}$ when going from $^{37}$K to $^{47}$K and then from $^{47}$K  to $^{49}$K. This result is very encouraging for these first-ever systematic {\it{ab initio}} calculations in mid-mass nuclei. Indeed, it allows one to speculate that correcting in the near future for the systematic overbinding produced in the Ca region by currently available chiral interactions, and thus for the too spread out spectra of odd-A systems, might bring the theoretical calculation in good agreement with experiment. Although this remains to be confirmed, it demonstrates that systematic spectroscopic data in mid-mass neutron-rich nuclei provide a good test case to validate/invalidate specific features of basic inter-nucleon interactions and innovative many-body theories.

To complement the above analysis, the lower panel of Fig.~\ref{AI-ESPE} provides the evolution of proton $1d_{3/2}$ and $2s_{1/2}$ shells. These two effective single-particle energies (ESPEs) recollect \cite{Soma2011} the fragmented $3/2^{+}$ and  $1/2^{+}$ strengths obtained from one-proton addition and removal processes on neighboring Ca isotones. Within the present theoretical description, the evolution of the observable (i.e. theoretical-scheme independent) lowest-lying $1/2^{+}$ and $3/2^{+}$ states does qualitatively reflect the evolution of the underlying non-observable (i.e. theoretical-scheme dependent) single-particle shells. As such, the energy gap between the two shells decreases from 5.76\,MeV in $^{39}$K to 1.81\,MeV in $^{47}$K which is a reduction of about $70$\%. Adding 4 neutrons in the $\nu 2p_{3/2}$ shell causes the energy difference to increase again to 4.03\,MeV.  
    
\subsection{Even-A}
The configuration of the even-A potassium isotopes arises from the coupling between an unpaired proton in the $sd$ shell with an unpaired neutron. Different neutron orbitals are involved, starting from $^{38}$K where a hole in the $\nu 1d_{3/2}$ is expected, then gradually filling the $\nu 1 f_{7/2}$ and finally, the $\nu 2p_{3/2}$ for $^{48,50}$K. 

In order to investigate the composition of the ground-state wave functions of the even-A K isotopes, we first compare the experimental magnetic moments to the semi-empirical values. Based on the additivity rule for the magnetic moments ($g$ factors) and assuming a weak coupling between the odd proton and the odd neutron, the semi-empirical magnetic moments can be calculated using the following formula \cite{Schwartz1953}:  $\mu_{\rm{se}}=g_{\rm{se}}\cdot I$, with

\begin{equation}
\label{eq:Empirical_g-factors}
\resizebox{.89\hsize}{!}{$g_{\rm{se}}=\frac{g(j_{\pi})+g(j_{\nu})}{2}+\frac{g(j_{\pi})-g(j_{\nu})}{2}\frac{j_{\pi}(j_{\pi}+1)-j_{\nu}(j_{\nu}+1)}{I(I+1)}$},
\end{equation}
where $g(j_{\pi})$ and $g(j_{\nu})$ are the experimental $g$ factors of nuclei with an odd proton or neutron in the corresponding orbital. The calculations were performed using the measured $g$ factors of the neighboring isotopes with the odd-even and even-odd number of particles in $j_{\pi}$ and $j_{\nu}$, respectively. For the empirical values of unpaired protons, results from Table\,\ref{mm-odd-A-Tab} were used. The $g$ factors for the odd neutrons were taken from the corresponding Ca isotones \cite{Minamisono1976,Andl1982,Olschewski1972,GarciaRuiz2013}. The obtained results with the list of isotopes used for different configurations are presented in Table~\ref{empirical_mm}. 
\begin{table}[h!]
\caption {Semi-empirical $g$ factors obtained for certain configurations using the additivity rule in Eq.\,(\ref{eq:Empirical_g-factors}) (see text for more details). In the calculations, results from Table\,\ref{mm-odd-A-Tab} were used for $g(j_{\pi})$, while for $g(j_{\nu})$ Ca data were taken from \cite{Minamisono1976,Andl1982,Olschewski1972,GarciaRuiz2013}. For $^{48}$K, different configurations are considered for $I=1$. \label{empirical_mm}}
\begin{ruledtabular}
\begin{tabular}{c c c e c}
Isotope & 
  $I^{\pi}$ & 
     configuration & 
        \multicolumn{1}{c}{$g_{\rm{se}}$} & 
           ($g(j_{\pi})$;$g(j_{\nu})$) \\
\hline
$^{38}$K & $3^{+}$ & $\pi 1d_{3/2}^{-1} \otimes \nu 1d_{3/2}^{-1}$ & +0.47 & ($^{39}$K; $^{39}$Ca ) \\ 
$^{40}$K & $4^{-}$ & $\pi 1d_{3/2}^{-1} \otimes \nu 1f_{7/2}$ & -0.31 & ($^{39}$K; $^{41}$Ca) \\ 
$^{42}$K & $2^{-}$ & $\pi 1d_{3/2}^{-1} \otimes \nu 1f_{7/2}^{3}$ & -0.64 & ($^{41}$K; $^{43}$Ca) \\ 
$^{44}$K & $2^{-}$ & $\pi 1d_{3/2}^{-1} \otimes \nu 1f_{7/2}^{5}$ & -0.62 & ($^{43}$K; $^{45}$Ca) \\ 
$^{46}$K & $2^{-}$ & $\pi 1d_{3/2}^{-1} \otimes \nu 1f_{7/2}^{-1}$ & -0.65 & ($^{45}$K; $^{47}$Ca) \\ 
$^{48}$K & $1^{-}$ & $\pi 1d_{3/2}^{-1} \otimes \nu 2p_{3/2}$ & -0.40 & ($^{45}$K; $^{49}$Ca) \\
$^{48}$K & $1^{-}$ & $\pi 2s_{1/2}^{-1} \otimes \nu 2p_{3/2}$ & -2.11 & ($^{47}$K; $^{49}$Ca) \\
\end{tabular}
\end{ruledtabular}
\end{table}

\begin{figure}[h!]
\includegraphics[width=0.95\linewidth]{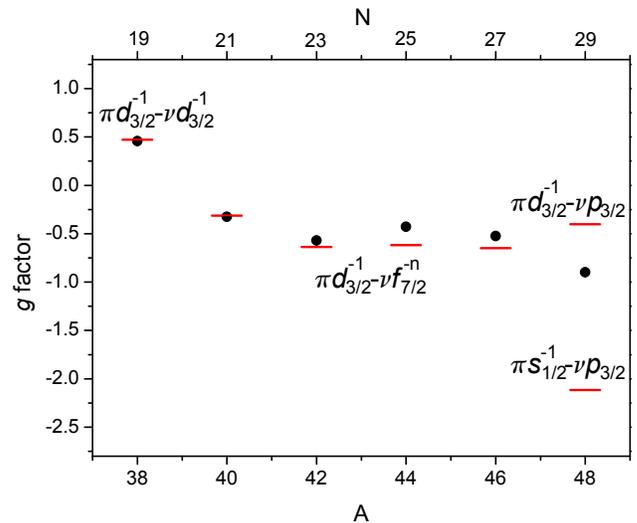}
\caption{\label{Even-A-emp} (color online) Experimental $g$ factors (black dots) compared to the semi-empirical values (red solid line) calculated from the neighboring isotopes. Based on the good agreement between the experimental and semi-empirical $g$ factors, the dominant component of the wave functions can be easily established for $^{38-46}$K. Only for $^{48}$K a strong mixing between the $\pi 2s_{1/2}^{-1} \otimes \nu 2p_{3/2}$ and the $\pi 1d_{3/2}^{-1} \otimes \nu 2p_{3/2}$ in the wave function is found.}
\end{figure}

A comparison between the experimental and semi-empirical $g$ factors is shown in Fig.~\ref{Even-A-emp}. 
For $^{38}$K, the semi-empirical value calculated from $^{39}$K and $^{39}$Ca provides excellent agreement with the experimental result. This confirms that the dominant component in the wave function for the ground state originates from the coupling between a hole in the $\pi 1d_{3/2}$ and the $\nu 1d_{3/2}$. By adding more neutrons, the $\nu 1f_{7/2}$ orbital is filled for $^{40}$K up to $^{46}$K. In order to calculate the semi-empirical $g$ factors for these isotopes, $g(j_{\pi})$ is provided from neighboring odd-A K isotopes (Table\,\ref{mm-odd-A-Tab}) combined with $g(j_\nu)$ of the subsequent odd-A Ca isotones starting from $N = 21$ up to $N = 27$. The trend of the experimental $g$ factors is very well reproduced by the semi-empirical calculations suggesting that the dominant component in the wave function of these isotopes is $\pi 1d_{3/2}^{-1} \otimes \nu 1f_{7/2}^{n}$ where $n=1,3,5,7$. 
For $^{48}$K, two semi-empirical values are calculated by considering a coupling between a proton hole in the $\pi 2s_{1/2}$ or the $\pi 1d_{3/2}$ with neutrons in the $\nu 2p_{3/2}$ orbital. Comparing the experimental $g$ factor to the semi-empirical results, it is possible to conclude that the main component in the wave function of this isotope arises from the configuration with a hole in the $\pi 1d_{3/2}$. Nevertheless, the deviation of the experimental result from the semi-empirical $g$ factors is due to a large amount of mixing between both configurations in the wave function. $^{50}$K is not presented because the observed $I = 0$ leads to $\mu = 0$. There are two possible configurations which would yield this particular spin: $\pi 1d_{3/2}^{-1} \otimes \nu 2p_{3/2}$ and $\pi 2s_{1/2}^{-1} \otimes \nu 2p_{1/2}$. 
 
\begin{figure}
\includegraphics[width=0.95\linewidth]{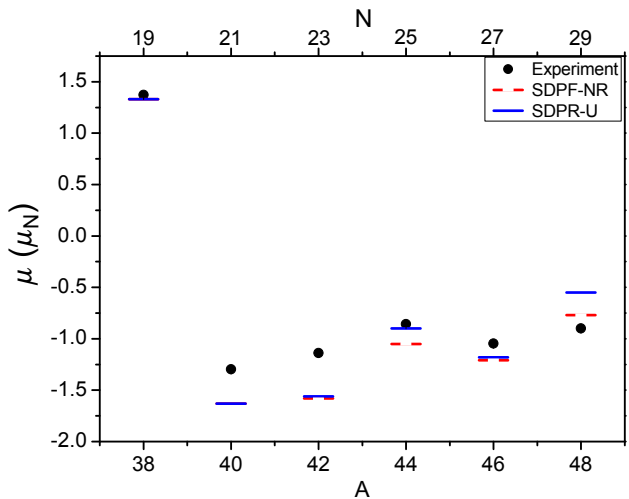}
\caption{\label{Even-A_mm_eff}(color online) Measured magnetic moments (black dots) for even-A K isotopes compared to the shell-model calculations using the SDPF-NR (red dashed line) as well as the SDPF-U (blue solid line) effective interaction. Although there is a larger deviation present for $^{40}$K and $^{42}$K, which might originate from lack of the excitations across $Z, N =20$, overall reasonable agreement between the experimental and theoretical results is observed. }
\end{figure}

The experimental magnetic moments together with shell-model calculations are summarized in Table~\ref{mm-even-A} 
\begin{table*}
\caption{Experimental magnetic moments (in units of $\mu_{\rm{N}}$) for even-A K isotopes compared to shell-model predictions using two effective interactions: SDPF-NR and SDPF-U. The error in the square brackets is due to the hyperfine structure anomaly, which amounts to 0.5\%. \label{mm-even-A}}
\begin{ruledtabular}
\begin{tabular}{c c z x x f c}
Isotope & 
  $I^{\pi}$ & 
    \multicolumn{1}{c}{$\mu_{\rm exp}$} & 
       \multicolumn{1}{c}{$\mu_{\rm{SDPF-NR}}$} & 
          \multicolumn{1}{c}{$\mu_{\rm{SDPF-U}}$} & 
             \multicolumn{1}{c}{$\mu_{\rm lit}$} & 
               Reference \\
\hline
$^{38}$K & $3^{+}$ & +1.3711\,(10)\,[69] & +1.33 & +1.33 & +1.371\,(6)\footnotemark[1] & \cite{Touchard1982} \\
$^{40}$K & $4^{-}$ & - & -1.63 & -1.63 & -1.2964\,(4)\footnotemark[2] & \cite{Eisinger1952}\\
$^{42}$K & $2^{-}$ & -1.1388\,(7)\,[57] & -1.58 & -1.56 & -1.14087\,(20)\footnotemark[2] & \cite{Chan1969}\\
$^{44}$K & $2^{-}$ & -0.8567\,(9)\,[43] & -1.05 & -0.90 & -0.856\,(4)\footnotemark[1] & \cite{Touchard1982}\\
$^{46}$K & $2^{-}$ & -1.0464\,(7)\,[52] & -1.21 & -1.18 & -1.051\,(6)\footnotemark[1] & \cite{Touchard1982}\\
$^{48}$K & $1^{-}$ & -0.8997\,(3)\,[45] & -0.77 & -0.55 & - & - \\
\end{tabular}
\end{ruledtabular}
\footnotetext[1]{Included 0.5\% uncertainty on the error to account for the hyperfine structure anomaly.}
\footnotetext[2]{The value without diamagnetic correction of +0.13\%. }
\end{table*}
and graphically presented in Fig.~\ref{Even-A_mm_eff}.
The predictions for $^{38}$K from both interactions reproduce the experimental magnetic moments very well. Furthermore, almost the same value is calculated with both interactions for $A=40$ and $A=42$, but the experimental results are underestimated by about 26\% and 37\% for $^{40}$K and $^{42}$K, respectively. While the SDPF-U interaction almost reproduces the observed magnetic moment for $^{44}$K, its earlier version (SDPF-NR) shows a deviation of approximately 0.26\,$\mu_{\rm{N}}$ when comparing to the experimental result. Additionally, very good agreement is observed between experimental and theoretical results for $^{46}$K. Finally, the situation is inverted for the case with the strongly mixed $^{48}$K, which is better reproduced by the SDPF-NR interaction and shows a deviation of about 0.35\,$\mu_{\rm{N}}$ for SDPF-U. The general trend of the magnetic moments is well reproduced by both interactions and the calculated magnetic moments are in reasonable agreement with the experimental results. The slightly larger deviation observed for $^{40}$K and $^{42}$K is probably due to lack of excitations across $Z, N = 20$.

At this point one should be aware that the odd-odd isotopes are more challenging for the shell-model calculation than odd-even nuclei due to the high level density at low energy. These levels arise from all different possibilities of couplings between an odd proton and an odd neutron. Although the energy of a calculated level might be wrong by hundreds of keV, if the magnetic moment is well-reproduced we can still draw reliable conclusions on the wave function composition of the state.   

\begin{figure}[h!]
\includegraphics[width=0.65\linewidth]{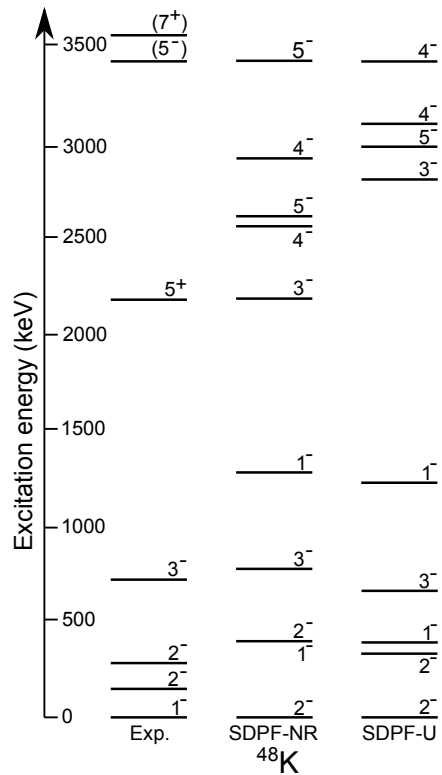}
\caption{\label{EL-48K} Experimental energy spectrum of $^{48}$K adapted from Ref.\,\cite{Krolas2011} using the fact that the nuclear spin is firmly established to be $1^{-}$ \cite{Kreim2014}. Results are compared to the calculated spectra from different effective interactions: SDPF-NR and SDPF-U.  }
\end{figure}

In the case of $^{38}$K, the $\pi 1d_{3/2}^{-1} \otimes \nu 1d_{3/2}^{-1}$ configuration constitutes more than 90\% of the total wave function. The dominant component of the ground-state wave function for all $N>20$ even-A K isotopes is arising from a hole in the $\pi 1d_{3/2}$ coupled to an odd neutron in the $pf$ orbital. For $^{40,42,44,46}$K, the main component is $\pi 1d_{3/2}^{-1} \otimes \nu 1f_{7/2}^{n}$ and its contribution to the wave function decreases from more than 90\% down to about 85\%. The lowest $1^-$ state in $^{48}$K is predicted to be an excited state by both interactions, respectively at $E = 407$\,keV and $E = 395$\,keV (see Fig.~\ref{EL-48K}). 
Both interactions favor a $2^{-}$ state as the ground state. In addition, an excited $2^-$ state is near-degenerate with the $1^-$ level, at $E=408$\,keV and $E=340$\,keV respectively. Considering the firmly assigned ground-state spin-parity of $^{48}$K, and using the multipolarities deduced from the measured lifetimes of the lowest four levels by Kr\'{o}las \textit{et al.} \cite{Krolas2011}, the experimental spin-parities of the four lowest excited states can now be more firmly assigned. A reasonable agreement with the calculated level scheme is shown up to 1\,MeV.  However, the positive parity level around 2\,MeV, which must be due to a proton excitation across the $Z=20$ gap, is not reproduced in the current calculations, as such excitations have not been included.

The wave function of the calculated lowest $1^-$ state, which reproduces the observed magnetic moment reasonably well, is very fragmented compared to the other even-A K isotopes: $\pi 1d_{3/2}^{-1} \otimes \nu 2p_{3/2}$ only constitutes approximately 40\% and 50\% of the total wave function for SDPF-NR and SDPF-U, respectively. The next leading component, $\pi 2s_{1/2}^{-1} \otimes \nu 2p_{3/2}$,  contributes only 15-20\%, although this isotope is located between two isotopes with a dominant $\pi 2s_{1/2}^{-1}$ configuration ($^{47}$K and $^{49}$K). In addition, configurations which arise from $1p1h$ excitation from $ \nu 1f_{7/2}$ to the rest of the $\nu (pf)$ shell have a significant contribution of about 15\% to the total wave function of the lowest $1^-$ state in $^{48}$K. In the case of $^{50}$K, the wave function of the $0^-$ level is much less fragmented: the main component is $\pi 1d_{3/2}^{-1} \otimes \nu (pf)$, constituting more than 85\% of the wave function. The contribution of the $\pi 2s_{1/2}^{-1}\otimes \nu (pf)$ component as well as the one from $1p1h$ neutron excitations is about 5\%. While this $0^-$ is correctly reproduced as the ground state by the SDPF-U interaction, it is predicted at 315\,keV (with a $2^-$ ground state) with SDPF-NR. 

\begin{figure}[h!]
\includegraphics[width=0.95\linewidth]{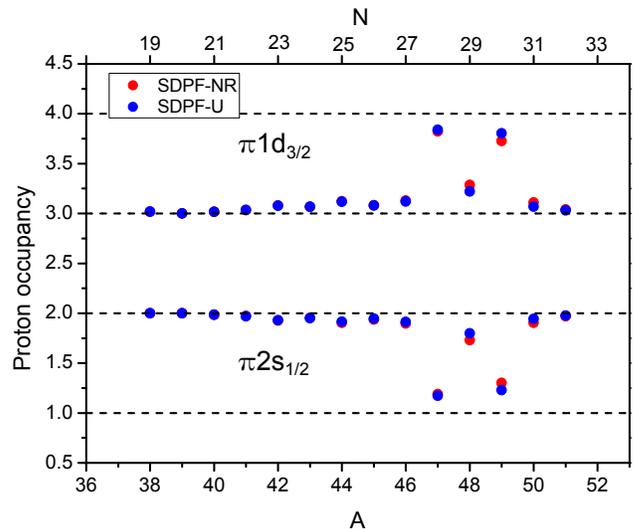}
\caption{\label{Occupation}(color online) Proton occupation of the $\pi 2s_{1/2}$ and the $\pi 1d_{3/2}$ orbitals from the shell-model calculations using the SDPF-NR and SDPF-U effective interactions. It is clear that for isotopes from $A = 38 - 46$ and $A =48, 50 - 51$ the dominant component in the configuration is a hole in the $\pi 1d_{3/2}$. In the case of $I^{\pi} = 1/2^{+}$ isotopes, a proton hole is located in the $2s_{1/2}$. Deviation from integer numbers for $^{47-49}$K indicates mixing in the wave function. }
\end{figure}

In addition to the magnetic moment and wave functions obtained from the shell-model calculations, it is also possible to extract information about the occupancy of the orbitals. The calculated occupancy of the $\pi 2s_{1/2}$ and $\pi 1d_{3/2}$ orbitals are shown in Fig.~\ref{Occupation}.
The maximum number of particles found in an orbital with total angular momentum $j$ is $2j+1$. Thus, for the $s_{1/2}$ this number is 2, while in case of the $d_{3/2}$ it is 4. The occupation of the $\pi 2s_{1/2}$ remains almost constant around 2 protons from $N=19$ up to $N=27$, with a slight decrease toward $^{46}$K. For these isotopes, the occupation of the $\pi 1d_{3/2}$ stays around 3 protons with a corresponding slight increase toward $A = 46$. This increase (decrease) of occupancy for the $\pi 1d_{3/2}$ ($\pi 2s_{1/2}$) orbital is probably due to the reduction of the energy difference between these two proton orbitals with increasing number of neutrons in the $\nu 1f_{7/2}$. Additionally, a small odd-even staggering in the proton occupation is also observed for these isotopes. This effect could be due to the proton-neutron coupling for the odd-odd isotopes, which results in a higher occupancy of $\pi 1d_{3/2}$ and a lower occupancy for the $\pi 2s_{1/2}$. In this region, there is no discrepancy observed between results from the different interactions. Furthermore, almost degenerate proton orbitals for $N = 28$ yield a hole in the $\pi 2s_{1/2}$ causing the $\pi 1d_{3/2}$ to be nearly completely filled. Surprisingly, for $^{48}$K with an additional unpaired neutron in the $\nu 2p_{3/2}$ orbital, the proton occupation of $\pi 1d_{3/2}$ drops down to about 3.3 protons while the $\pi 2s_{1/2}$ occupation increases accordingly. For the next isotope with two neutrons placed in the $\nu 2p_{3/2}$ ($^{49}$K), the occupation of the proton orbitals is more similar to $^{47}$K, where a hole in the $\pi 2s_{1/2}$ is found. This is also in agreement with the nuclear spins and magnetic moments of these two isotopes. At this point a larger deviation from integer numbers for the proton occupation indicates a larger amount of mixing in the configurations of $^{47\text{-}49}$K. Based on the information obtained from the $g$ factor and magnetic moment for $^{48}$K, a hole in the $\pi d_{3/2}$ was expected  to be the dominant component, which is confirmed by these occupancies. Nevertheless, the reason for the big decrease of the $\pi 1d_{3/2}$ occupancy compared to the neighboring two isotopes is still puzzling. Adding one and two more neutrons leads to the "normal" occupation for the neutron-rich K isotopes with the filled $\pi 2s_{1/2}$ and a hole in the $\pi 1d_{3/2}$.     

\section{Summary}

Hyperfine spectra of potassium isotopes between $N = 19$ and $N = 32$ were measured using collinear laser spectroscopy, yielding the nuclear spins and magnetic moments. The experimental results were compared to shell-model calculations using two different effective interactions: SDPF-NR and SDPF-U. Overall good agreement is observed between the measured magnetic moments and theoretical predictions. This allows one to draw conclusions on the composition of the wave function as well as on the proton occupation of the $2s_{1/2}$ and $1d_{3/2}$ orbitals. It was shown that the dominant component of the ground-state wave function for odd-A isotopes up to $^{45}$K arises from a hole in the $\pi 1d_{3/2}$. Additionally, for isotopes with spin $1/2^{+}$ the main component of the wave function is $\pi 2s_{1/2}^{-1}$ with more mixing present in $^{49}$K coming from the almost degenerate $\pi 2s_{1/2}$ and $\pi 1d_{3/2}$ proton orbitals. The nuclear spin of $^{51}$K, which was found to be $3/2$, points to the "normal" ordering of the EPSE, and this is confirmed by the measured magnetic moment that is close to the $\pi 1d_{3/2}$ single particle value. In the case of odd-odd isotopes, the main configuration originates from the coupling of the $\pi 1d_{3/2}^{-1} \otimes \nu (pf)$, and this for all odd-odd isotopes from $N=19$ up to $N = 31$. Only for $^{48}$K, a very fragmented wave function has been observed for the $1^-$ ground state. This level becomes the ground state due to a significant ($>$20\%) contribution from the $\pi 2s_{1/2}^{-1} \otimes \nu (pf)$ configuration. Moreover, a detailed discussion about the evolution of the proton effective single particle energies (ESPEs) was presented. The central term of the monopole interaction was found to have the strongest effect in the changing ESPE beyond $N = 28$. {\it{Ab initio}} calculations of the ESPEs show a considerable decrease (70\%) of the gap between $\pi (1d_{3/2}-2s_{1/2})$ at $N = 28$ presenting a promising starting point for the approach which is currently still under development. 
 
The experimental results of the neutron-rich potassium isotopes have a relevant role in the future improvements of the effective shell-model interactions and \textit{ab initio} calculations. Additional experimental data for $^{51}$K and $^{53}$K, in particular the spin of the $^{53}$K ground state and the energy of the $I = 1/2 ^{+}$ states, could provide the final clues about the evolution of the proton $sd$ levels in this region.

\begin{acknowledgments}
The authors are grateful to F. Nowacki for providing the effective interactions which were used in the shell-model calculations. C.B., T.D. and V.S. would like to thank A. Cipollone and P. Navr\'{a}til for their collaboration on $3N$  forces. This work was supported by the IAP-project P7/12, the FWO-Vlaanderen, GOA 10/010 from KU Leuven, NSF Grant PHY-1068217, BMBF (05 P12 RDCIC), Max-Planck Society, EU FP7 via ENSAR (No. 262010) and STFC Grant No. ST/L005743/1. Gorkov-Green's function calculations were performed using HPC resources from GENCI-CCRT (Grant No. 2014-050707) and the DiRAC Data Analytic system at the University of Cambridge (BIS National E-infrastructure Capital Grant No. ST/K001590/1 and STFC Grants No. ST/H008861/1, ST/H00887X/1, and ST/K00333X/1). We would like to thank to the ISOLDE technical group for their support and assistance.
\end{acknowledgments}

\bibliography{References-short}

\end{document}